\documentclass[runningheads]{llncs}

\usepackage{amsmath}
\usepackage{amsfonts}
\usepackage{amssymb}
\usepackage{booktabs}
\usepackage{graphicx}
\usepackage[colorlinks]{hyperref}
\usepackage[switch]{lineno}
\usepackage[dvipsnames]{xcolor}
\usepackage{subcaption}
\usepackage{bm}

\newcommand\blfootnote[1]{%
  \begingroup
  \renewcommand\thefootnote{}\footnote{#1}%
  \addtocounter{footnote}{-1}%
  \endgroup
}

\usepackage[T1]{fontenc}

\usepackage{color}

\newcommand{\winners}{\mathcal{W}}
\newcommand{\losers}{\mathcal{L}}
\newcommand{\cand}{\mathcal{C}}

\newcommand{\assertion}{\mathcal{A}}
\newcommand{\quota}{Q}
\newcommand{\seats}{N}
\newcommand{\ballots}{\mathcal{B}}
\newcommand{\election}{E}

\newcommand{\IQXb}{\textsf{\textbf{IQX}}}

\newcommand{\IQb}{\textsf{\textbf{IQ}}}
\newcommand{\UTb}{\textsf{\textbf{UT}}}
\newcommand{\LTb}{\textsf{\textbf{LT}}}

\newcommand{\IQX}{\textsf{IQX}}
\newcommand{\AG}{\textsf{AG}}
\newcommand{\IQ}{\textsf{IQ}}
\newcommand{\UT}{\textsf{UT}}
\newcommand{\LT}{\textsf{LT}}

\newcommand{\CNEBS}{\textsf{NL}^*}
\newcommand{\CNEBSb}{\textsf{\textbf{NL}}^*}
\DeclareMathOperator{\first}{first}

\newcommand{\proj}{\sigma}
\newcommand{\AGS}{\textsf{AG}^*}
\newcommand{\AGSb}{\textsf{\textbf{AG}}^*}

\newcommand{\ignore}[1]{}

\newcommand{\mysubsection}[1]{\paragraph{\rm \textbf{#1.}}}

\title{3+ Seat Risk-Limiting Audits for Single Transferable Vote Elections}

\author{
Michelle Blom     \inst{1}   \orcidID{0000-0002-0459-9917}  \and
Alexander Ek \inst{2} \orcidID{0000-0002-8744-4805}
\and
Peter J. Stuckey  \inst{3}   \orcidID{0000-0003-2186-0459}  \and
Vanessa Teague    \inst{4}   \orcidID{0000-0003-2648-2565}  \and
Damjan Vukcevic   \inst{2} \orcidID{0000-0001-7780-9586}}

\authorrunning{M.~Blom et al.}

\institute{
School of Computing and Information Systems, University of Melbourne,
Parkville, Australia \\
\email{michelle.blom@unimelb.edu.au}
\and
Department of Econometrics and Business Statistics, Monash University, Clayton,
Australia
\and 
Department of Data Science and AI, Monash University, Clayton, Australia
\and
Thinking Cybersecurity Pty.\ Ltd., Melbourne, Australia
}

\begin{document}

\maketitle

\begin{abstract}
Constructing efficient risk-limiting audits (RLAs) for multi-winner single transferable vote (STV) elections is a challenging problem.
An STV RLA is designed to statistically verify that the reported winners of an election did indeed win according to the voters' expressed preferences and not due to mistabulation or interference, while limiting the risk of accepting an incorrect outcome to a desired threshold (the risk limit).
Existing methods have shown that it is possible to form RLAs for two-seat STV elections in the context where the first seat has been awarded to a candidate in the first round of tabulation.
This is called the \textit{first winner criterion}. 
We present an assertion-based approach to conducting full or partial RLAs for STV elections with three or more seats, in which the first winner criterion is satisfied.
Although the chance of forming a full audit that verifies all winners drops substantially as the number of seats increases, we show that we can quite often form partial audits that verify most, and sometimes all, of the reported winners. 
We evaluate our method on a dataset of over 500 three- and four-seat STV elections from the 2017 and 2022 local council elections in Scotland.\blfootnote{To appear at the 10$^{th}$ Workshop on Advances in in Secure Electronic Voting Schemes.}  
\end{abstract}

\section{Introduction}

The single transferable vote (STV) is a multi-winner proportional ranked-choice election system. Voters cast a ballot in which they rank available candidates in order of preference. Depending on the jurisdiction, votes may be partial, expressing a ranking over only a subset of candidates, or total, in which all candidates are ranked. In general, candidates are awarded a seat when they amass more than a certain share of the total vote, called a \textit{quota}. This quota will depend on the number of available seats and the total number of valid ballots cast. Tabulation is complex, as while each cast ballot starts with a value of 1, this value changes as tabulation proceeds. This complexity makes auditing STV elections hard. In this paper, we build upon earlier work on 2-seat STV risk-limiting audits (RLAs)  \cite{blom2022stvrlasSHORT,blom2024stvrlasSHORT}, presenting  a generalisation of the prior method to $\seats$-seat RLAs.

STV tabulation proceeds in rounds of candidate election and elimination. Initially, each candidate is given all of the ballots on which they have been ranked first. Candidates are elected when their tally reaches or exceeds the election's quota. Where multiple candidates reach a quota in the same round, they are elected in order of their tallies, highest first. The ballots in a seated candidates' tally pile are subsequently re-weighted, and given to the next most preferred eligible candidate in their rankings. This re-weighting is designed to remove a quota's worth of votes from the system, and can be accomplished in myriad ways. In this paper, we consider the variant of STV that uses the Weighted Inclusive Gregory method. The current value of each ballot in the winner's tally pile is multiplied by a \textit{transfer value} equal to the difference between their total tally and the quota (their \textit{surplus}) divided by their total tally. In any round where no candidate has a quota, the candidate with the smallest tally is eliminated, with their ballots given, at their current value, to candidates preferenced further down the list. This variant of STV is used in the United States and Scotland.

Blom et. al. \cite{blom2022stvrlasSHORT} presented a first approach for constructing an assertion-based STV RLA in the 2-seat context. Two cases were considered separately: (i) where a candidate was awarded a seat in the first round of tabulation, on the basis of their first preference tally (the \textit{first winner} criterion), and (ii) where a candidate was eliminated in the first round. Case (i) was found to be more straightforward to audit. The audit consisted of assertions to verify that: the candidate who won in the first found did indeed have a quota's worth of votes, and that the second winner could not possibly lose to all the reported losers. A method designed to tackle the second case was presented, but found to be impractical in general.  

In follow-up work, Blom et. al. \cite{blom2024stvrlasSHORT} showed how 2-seat STV RLAs, where the first winner criterion was satisfied, could be made more efficient by substantially reducing the number of ballots to be sampled in the audit. Previously, assertions were formed to establish an upper bound on the first winner's transfer value. This upper bound was then used to help form assertions to show that the second winner beat all of the reported losers. These assertions compared the minimum possible tally of the second winner against the maximum possible tally of the reported losers. This kind of assertion could only be formed when the minimum tally of the winner was greater than the maximum tally of the loser. The transfer value upper bound was used to reduce these maximum tallies. Blom et. al. \cite{blom2024stvrlasSHORT} extended this earlier work by adding assertions to establish a lower bound on the first winner's transfer value. This bound was used to increase the winner's minimum possible tally in these minimum-maximum tally comparisons. The result was both assertions that are more efficient to audit, and an increase in the number of assertions that could be formed. This work additionally introduced the concept of a \textit{partial} audit, where all winners were not verified, but some candidates are shown to have definitely lost, and some to have definitely won. 

The approach we present in this paper is restricted to the context where \textit{at least one candidate has been seated in the first round} of tabulation. Our method forms either a full RLA, where all winners can be verified, or a partial RLA, in which we verify only a subset. One new assertion type is presented to help form these audits.  We evaluate this approach on a data set of 513 three- or four-seat Scottish local council STV elections, all satisfying the first winner criterion. Of the 252 three-seat contests, we can verify all winners in  58\% of instances, two of the three winners in 29\%, one winner in 11\%, and no winners in 2\%. Of the 261 four-seat contests, we verify all four winners in 32\% of instances, three winners in 30\%, two winners in 26\%, one winner in 10\%, and no winners in 2\%.

\vspace{-5pt}
\section{Single Transferable Vote}

 STV tabulation proceeds in rounds of candidate election and elimination.  As candidates are elected or eliminated, the ballots in their tally piles are re-distributed to the next most preferred \textit{eligible} candidate.  Candidates are eligible to receive votes in a round iff: they have not yet been elected or eliminated, and they did not already have a quota at the start of the round.\footnote{If a candidate reaches a quota mid-transfer, the transfer does not stop. The candidate will continue to receive all votes in the transfer designated for them. }  Tabulation stops when  all seats have been filled, or the number of candidates remaining equals the number of unfilled seats, at which point all remaining candidates are elected to a seat.

\autoref{tab:EGSTV1} presents an example  3-seat STV election with five candidates, $A$ to $E$, tabulated using the Weighted Inclusive Gregory Method. The quota is 308 votes.  The first preference tallies of $A$ to $E$ are 250, 120, 400, 350 and 110 votes, respectively. Candidates $C$ and $E$ have a quota on first preferences. Candidate $C$ has the largest surplus, at 202 votes, and is elected first. Their transfer value is $\tau_1 = 208/510 = 0.396$. The 400 ${}[C,D]$ ballots are each re-weighted to 0.396, and 158.4 votes are added to $D$'s tally. The 110 ${}[C,E,D]$ ballots are also re-valued to 0.396, and are given to $D$, skipping $E$ as $E$ already has a quota. 
Candidate $E$ is then elected. Their transfer value would be $\tau_2 = 42/350 = 0.12$, but all of the ballots in their tally pile exhaust, with no later eligible preferences. In the third round, no candidate has a quota's worth of votes, and the candidate with the smallest tally, $B$, is eliminated. The 120 ${}[B,A,C]$ ballots go to $A$, each retaining their current value of 1. At the start of the fourth round, candidate $A$ has reached a quota, at 370 votes, and is elected to the third and final seat.

\begin{table}[t]
    \begin{subtable}{.3\columnwidth}
      \centering
        \begin{tabular}{lr}
& \\
& \\
\hline
Ranking & Count \\
\hline
{}[$A$] & 250 \\
{}[$B$,$A$,$C$] & 120 \\
{}[$C$,$D$] & 400 \\
{}[$E$] & 350 \\
{}[$C$,$E$,$D$] & 110  \\
\hline
\end{tabular}
        \caption{}
				\label{tab:EGSTV1a}
    \end{subtable}
    \begin{subtable}{.7\columnwidth}
      \centering
        \begin{tabular}{crrrr}
$\seats$: 3 & $\quota$: 308  & & &\\
\hline
Cand. & Round 1 & Round 2 & Round 3 & Round 4\\
\hline
      & $C$ elected      & $E$ elected           & $B$ elim. & $A$ elected\\
      & $\tau_1 = 0.396$ & $\tau_2 = 0.12$  &  & \\
\hline
$A$ & 250  & 250 & 250 & \textbf{370}\\
$B$ & 120  & 120 & 120 & --\\
$C$ & \textbf{510}  & -- & -- & --\\
$D$ & 0  & 201.96 &  201.96 & 201.96\\
$E$ & \textbf{350}  & \textbf{350} & -- & --\\
\hline
\end{tabular}
\caption{}
\label{tab:EGSTV1b}
    \end{subtable} 
    \caption{3-seat STV election, quota 308 votes, stating (a) the number of
ballots cast with each listed ranking over candidates $A$ to $E$, and (b)
the tallies for each candidate after each round, and when a quota is reached (in bold). \vspace{-20pt}}
\label{tab:EGSTV1}
\end{table}

\vspace{-5pt}
\section{Assertion-based RLAs}

An assertion-based RLA is an RLA in which we statistically test a set of statements about a contest, called assertions. An \emph{assertion} is a statement about the full set of ballots cast in an election, typically expressed as an inequality  comparing the number of votes in one category against the number of votes in another. To construct an audit in the SHANGRLA framework \cite{shangrlaSHORT}, we need to design a set of assertions such that, if they are all true, they imply that the reported winner(s) really won the election. 
In general, any linear combination of tallies (counts of different types of ballots) can be converted into a SHANGRLA assertion \cite{blom2021assertionSHORT}.

\section{Preliminaries and Notation}

We define an STV election as per prior work \cite{blom2022stvrlasSHORT,blom2024stvrlasSHORT}. A ballot $b$ is  a sequence of candidates $\pi$, listed in order of preference (most popular first), without duplicates but without necessarily including all candidates. We use list
notation (e.g., $\pi = [c_1,c_2,c_3,c_4]$) and $\first(\pi) = \pi(1)$
to denote the first candidate in  sequence $\pi$. 

\begin{definition}[STV Election]\label{def:STV}
An STV election $\election$ is a tuple $\election = (\cand, \ballots, \quota,
\seats)$ where $\cand$ is a set of candidates,  $\ballots$ the multiset of
ballots cast,\footnote{A multiset allows for the
inclusion of duplicate items.} $\quota$ the election quota (the number of votes a candidate must
attain to win a seat---usually the Droop quota---\autoref{eqn:Droop}), and
$\seats$ the number of seats to be filled.
\begin{equation}
\quota = \left\lfloor \frac{|\ballots|}{\seats + 1} \right\rfloor + 1
\label{eqn:Droop}
\end{equation}
\end{definition}

We use the concept of projection of one set onto another to reason about who a ballot belongs to when we assume that only a subset of candidates are `still standing' (i.e., they have not yet been eliminated or elected to a seat). 

\begin{definition}[Projection $\mathbf{\proj_\mathcal{S}(\pi)}$]
We define the projection of a sequence $\pi$ onto a set $\mathcal{S}$ as the
result of filtering all elements from $\pi$ that are not in $\mathcal{S}$. (The
elements keep their relative order in $\pi$.) For example:
$\proj_{\{c_2,c_3\}}([c_1,c_2,c_4,c_3]) = [c_2,c_3]$ 
 and $\proj_{\{c_2,c_3,c_4,c_5\}}([c_6,c_4,c_7,c_2,c_1]) = [c_4,c_2].$
\label{def:Projection}
\end{definition}

We use notation $t_{c,r}$ to denote the \emph{tally} of candidate $c$ in round $r$. The tally of each candidate in the first round is comprised of all the ballots on which they are ranked first (also referred to as their first preference tally). 
Note that, unlike in most elections, tallies are not necessarily integers.

The cost of auditing a given assertion refers to the estimated number of ballots we expect will need to be sampled in order to verify it--it's \emph{approximate sample number} (ASN). The estimated cost of an audit as a whole, consisting of a set of assertions, is the maximum of the expected costs of its assertions. 

We use $\assertion$ to denote a set of assertions, $a$ to denote a single assertion, $ASN(a)$ to denote the expected sample size required to audit $a$, and $ASN(\assertion)$ to denote the maximum ASN across all assertions in $\assertion$. We use the notation $M$ to denote the maximum sample size that we would consider to be \textit{auditable}. Throughout this paper, we consider an assertion $a$ to be \textit{auditable} if its ASN is less than or equal to a defined threshold, $M$ (i.e., $ASN(a) \leq M$). 

For an $N$-seat STV election $\election$ using the Droop quota (\autoref{eqn:Droop}), the maximum theoretical transfer value computable for any winning candidate is known to be $\tau_{max} = N/(N+1)$. For a three-seat contest, for example, $\tau_{max} = 0.75$. The maximum transfer value arises when a candidate achieves the maximum possible vote (i.e., all the votes). It is clear that this is no more than $(N+1) \quota$. This gives a transfer value of $\tau_{max} = ((N+1)\quota - \quota)/((N+1)\quota)) = N/(N+1)$. 

\section{Assertions}\label{sec:Assertions}

\paragraph{Existing Assertions}\label{sec:ExistingAssertions}

The following five assertion types are re-used from prior work \cite{blom2022stvrlasSHORT,blom2024stvrlasSHORT}, although with some slight changes, as noted.

\noindent
$\IQb(c)$.
Candidate $c$'s first-preference tally is at least a quota: $t_{c,1} \geq \mathcal{Q}$.
\vspace{4pt}

\noindent In the context where candidate $c$ has been elected on their first preferences:

\noindent $\UTb(c, \overline{\tau}_c)$.
Candidate $c$'s transfer value is less than $\overline{\tau}_c$: $t_{c,1} < \quota / (1 - \overline{\tau}_c)$.
\\ \noindent 
$\LTb(c, \underline{\tau}_{c})$.
Candidate $c$'s transfer value is greater than $\underline{\tau}_{c}$: $t_{c,1} > \quota / (1 - \underline{\tau}_{c})$.
\vspace{4pt}

\noindent In the context where candidates $W$ have been \textit{elected on their first preferences} with lower and upper bounds on their transfer values,  $\bm{\underline{\tau}}$ and $\bm{\overline{\tau}}$:

\noindent $\AGSb(w, l, W, \bm{\underline{\tau}}, \bm{\overline{\tau}})$.  The minimum tally of candidate $w$ is greater than the maximum tally of $l$. Thus,  $w$ will always have higher tally than $l$.  The contributions of ballot $b \in \ballots$ to $w's$ minimum tally, and $l$'s maximum tally, are:
\begin{flalign*}
C^{\AGS}_{min}(b, w, W, \bm{\underline{\tau}}, \bm{\overline{\tau}}) & = \left\{
\begin{array}{ll}
1  & \first(b) = w \\
\underline{\tau}_{\first(b)}\,\, & \first(\proj_{\cand - W}(b)) = w  \\
0 & \text{otherwise}
\end{array}
\right. \\
C^{\AGS}_{max}(b, l, W, \bm{\underline{\tau}}, \bm{\overline{\tau}}) & = 
\left\{
\begin{array}{ll}
0 & l \text{ does not occur in } b \\
0 & w \text{ appears before } l \text{ in } b \\
\overline{\tau}_{\first(b)} & \first(b) \in W  \\
1 & \text{otherwise} \\
\end{array}
\right.
\end{flalign*}
We say that $\AGS(w, l, W, \bm{\underline{\tau}}, \bm{\overline{\tau}})$ iff $t1^{min}_w > t1^{max}_l$, where
\[
t1^{min}_w  = \sum_{b \in \ballots} C^{\AGS}_{min}(b, w, W, \bm{\underline{\tau}}, \bm{\overline{\tau}})  \quad \text{and} \quad
t1^{max}_l  = \sum_{b \in \ballots} C^{\AGS}_{max}(b, l, W, \bm{\underline{\tau}}, \bm{\overline{\tau}})  
\]

\noindent $\CNEBSb(w,$ $l,W,$ $\bm{\underline{\tau}},\bm{\overline{\tau}},O^*)$. Candidate $w$ will always have a higher tally than $l$ given that  $O^*$ is the set of candidates $o \in \cand$ for which $\AGS(w, o, W, \bm{\underline{\tau}}, \bm{\overline{\tau}})$ holds. The assertion holds if the minimum tally of $w$, in this context, is greater than the maximum of $l$.  We define $w$'s minimum tally at a point at which they could be eliminated, where $O^*$ must already be eliminated. The contribution of $b \in \ballots$ to the minimum tally of $w$, and the maximum tally of $l$, is:
\begin{flalign*}
C^{\CNEBS}_{min}(b, w, W, \bm{\underline{\tau}}, \bm{\overline{\tau}}, O^*) & = \left\{
\begin{array}{ll}
1  & \first(\proj_{\cand - O^*}(b)) = w \\
\underline{\tau}_{\first(b)}\,\, & \first(\proj_{\cand - W}(b)) = w    \\
0 & \text{otherwise}
\end{array}
\right. \\
C^{\CNEBS}_{max}(b, l, W, \bm{\underline{\tau}}, \bm{\overline{\tau}}) & = \left\{
\begin{array}{ll}
0 & l \text{ does not occur in } b \\
0 & w \text{ appears before } l \text{ in } b \\
\overline{\tau}_{\first(b)}\,\, & \first(b) \in W    \\
1 & \text{otherwise} \\
\end{array}
\right.
\end{flalign*}
We say that $\CNEBS(w, l, W, \bm{\underline{\tau}}, \bm{\overline{\tau}}, O^*)$ iff $t2^{min}_w > t2^{max}_l$, where:
\[
t2^{min}_w  = \sum_{b \in \ballots} C^{\CNEBS}_{min}(b, w, W, \bm{\underline{\tau}}, \bm{\overline{\tau}}, O^*)  \quad \text{and} \quad 
t2^{max}_l  = \sum_{b \in \ballots} C^{\CNEBS}_{max}(b, l, W, \bm{\underline{\tau}}, \bm{\overline{\tau}})  
\]

 In earlier work \cite{blom2024stvrlasSHORT}, $\AGS$'s and $\CNEBS$'s were defined in the context where we assume $W$ have already been seated, but not necessarily on their first preferences. In this paper, we use the stricter assumption that $W$ were seated on first preferences as (i) this is the only context in which we form the assertions, and (ii) the stricter assumption allows us to provide more straightforward definitions of how each ballot contributes to the minimum and maximum tallies of candidates.

\paragraph{New Assertions}

With $W$, $\bm{\underline{\tau}}$ and $\bm{\overline{\tau}}$ defined as above:

\noindent
$\IQXb(w, W, \bm{\underline{\tau}}, \bm{\overline{\tau}}, O^*)$.
The total of candidate $w$'s first preference tally, in addition to any votes that would flow to them upon the election of candidates in $W$ and elimination of all candidates $O^*$, where $O^*$ is the set of candidates $o \in \cand$ for which $\AGS(w, o, W, \bm{\underline{\tau}}, \bm{\overline{\tau}})$ holds, is at least a quota. If true, $w$ is guaranteed to be awarded a seat. We define $w$'s $\IQX$ tally as:
\begin{equation*}
T_w^\IQX = \sum_{b \in \ballots} C^\IQX(b, w, W, \bm{\underline{\tau}}, \bm{\overline{\tau}}, O^*)
\end{equation*}
\begin{flalign*}
C^{\IQX}(b, w, W, \bm{\underline{\tau}}, \bm{\overline{\tau}}) & = \left\{
\begin{array}{ll}
1 & \first(\proj_{\cand - O^*}(b)) = w \\
\underline{\tau}_{\first(b)}\,\, & \first(b) \in W  \text{ and }  \first(\proj_{\cand - W}(b)) = w \\
0 & \text{otherwise} \\
\end{array}
\right.
\end{flalign*}

\section{Overview: Two Audit Options}\label{sec:Overview}

Our approach considers two alternate ways of forming an RLA for a 3+ seat STV election: an audit that consists only of $\IQX$ assertions; and an audit formed by a dual-loop heuristic.  
In this section, we describe the straight $\IQX$ audit and give an overview of the dual-loop heuristic (presented in greater detail in  \autoref{sec:Detail}).

Consider an STV election  $\election = (\cand, \ballots, \quota,
\seats)$ with reported winners $\winners$ and losers $\losers$. 
Let $\winners_0$ denote the subset of reported winners  that, in the reported election outcome, have won on first preferences. This is distinct from the notation $W$, which in the context of an assertion denotes the subset of candidates that we \textit{assume} have been seated on first preferences ($W \subseteq \winners_0$). 

Let $\assertion_{E,\IQX}$ and $\assertion_{E,DL}$ denote the set of assertions forming, respectively, a straight $\IQX$ audit  
and a dual-loop audit. 
If we can form a full audit via either approach,  we choose the audit with the least expected cost, and  the best partial audit formed by the dual-loop method, otherwise.

\mysubsection{Option 1: Straight $\IQX$ Audit}\label{sec:StraightIQX}
We consider a context where we \textit{do not} make assumptions about who has won on first preferences, and compute the set of all auditable $\AGS$ relationships we can form between reported winners and losers.
\[
\assertion_{\AG} \leftarrow [\AGS(w, l, W=\emptyset, \bm{\underline{\tau}}=\emptyset, \bm{\overline{\tau}}=\emptyset) | \forall w \in \winners, l \in \losers]
\]

We then compute the set of all auditable $\IQX$ relationships we can form for reported winners in this same context, where, for a given $w \in \winners$, $O^*$ is the set of candidates $o \in \cand$ for which $\AGS(w, o, \emptyset,$ $\emptyset, \emptyset)$  $\in \assertion_{\AG}$.
\[
 \assertion_{\IQX} \leftarrow [\IQX(w, W=\emptyset, \bm{\underline{\tau}=\emptyset}, \bm{\overline{\tau}}=\emptyset, O^*) | \forall w \in \winners]
\]

 If we can form an auditable $\IQX$ assertion for each reported winner, we can form a complete audit with just these assertions and any $\AGS$ relationships used to form them. Let $\assertion_{E,\IQX}$ denote this set of assertions.

\mysubsection{Option 2: Dual-Loop Audit} \label{sec:DualLoop}
If the straight $\IQX$ audit is not possible ($\assertion_{E,\IQX} = \emptyset$), or we want to try and find an audit with a smaller sample size, we shift our context to one where we do assume that at least one of our reported winners have won on their first preferences. \autoref{fig:overviewdualloop} provides an outline of the `outer' loop of the dual-loop method. We describe this algorithm in detail, below.

We first create the set of all auditable $\IQ$ assertions we can form for winners in $\winners_0$ (step 1), forming the set $\assertion_{\IQ}$. If $\assertion_{\IQ} = \emptyset$ then we report that an audit is not possible by the dual-loop method (step 4). We need to verify at least one reported winner with an $\IQ$ assertion to proceed with this second audit type.

If $|\assertion_{\IQ}| = |\winners|$ then we use this set of assertions, $\assertion_{E,DL} = \assertion_{\IQ}$, to form our audit (step 5). If $|\assertion_{\IQ}| < |\winners|$, we have a set of reported winners $W' \subseteq \winners_0$ whose win on first preferences we can verify, and a set of other winners $\mathcal{R} = \winners \setminus W'$ (steps 6--7). We then need to verify that each $r \in \mathcal{R}$ deserved to win.

This is accomplished with a heuristic, similar to that of \cite{blom2024stvrlasSHORT}, with an outer and inner loop. The outer loop performs a neighbourhood search over transfer value lower bounds for the candidates in $W'$, while the inner loop performs a neighbourhood search over transfer value upper bounds.  In the 3+ seat context, these loops will need to search through spaces of upper and lower bound vectors.   For a given pair of lower and upper bound vectors, our heuristic attempts to find a candidate full audit that verifies each $w \in \winners$, and if this is not possible, establish a candidate partial audit that verifies some but not all of these winners. The heuristic keeps track of the best (cheapest) full audit we can form, and the partial audit that verifies the most winners with the least expected cost.

\vspace{.3em}
\noindent \textbf{Outer Loop.} Given a set of verified winners on first preferences, $W'$, and the $\IQ$ assertions used to verify them, the loop searches for the best full and partial audit that it can find (steps 9--21 of \autoref{fig:overviewdualloop}). It  performs a neighbourhood search over candidate transfer value lower bound vectors, starting from 0 and gradually increasing these bounds in small increments $\delta \ll 1$, executing an inner loop for each of these vectors (step 14), and replacing $\assertion_{full}$ and $\assertion_{partial}$ when a better full and partial audit is found (steps 15--18).

\vspace{.3em}
\noindent \textbf{Inner Loop.}  Given a vector of lower bounds on first winner transfer values,  $\bm{\underline{\tau}}$, performs a neighbourhood search over the space of transfer value upper bound vectors (\autoref{alg:innerloop}), starting their reported values and gradually increasing these bounds in small increments $\delta \ll 1$.  For each candidate  $\bm{\overline{\tau}}$, together with $\bm{\underline{\tau}}$, a procedure \textsc{ConstructAudit} returns a candidate full audit, $\assertion'_{full}$, and partial audit, $\assertion'_{partial}$. We use the same increment size $\delta$ for both  loops, however different parameters could be used. A smaller  $\delta$ can allow us to find cheaper audits, at the cost of increasing the heuristic's run time. Note that the ordering of the two loops is arbitrary, the heuristic could be equivalently re-framed with the outer loop searching over upper bounds, and the inner  over lower bounds.

\vspace{.3em}
\noindent \textbf{Audit Formation.}  A procedure \textsc{ConstructAudit} that, given a pair of lower and upper bound vectors on first winner transfer values, $\bm{\underline{\tau}}$ and $\bm{\overline{\tau}}$, returns a candidate full audit, $\assertion'_{full}$, and partial audit, $\assertion'_{partial}$ (\autoref{fig:constructaudit}). If the procedure could not form all the assertions necessary to verify both the validity of the given lower and upper bounds, and each reported winner, then $\assertion'_{full} = \emptyset$ and $ASN(\assertion'_{full}) = \infty$. The partial audit will, at least, verify the set of first winners in $W'$, with the assertion set $\assertion_{\IQ}$. If $\assertion'_{full} \neq \emptyset$, then $\assertion'_{partial} \equiv \assertion'_{full}$.

\begin{figure}[!tp]
\centering
\begin{tabbing}
\hspace{0.25in}\=\hspace{0.25in}\=\hspace{0.25in}\=\hspace{0.25in}\=\hspace{0.25in}\=\hspace{0.25in}\=\kill
\textsc{DualLoopAudit}($\election = (\cand, \ballots, \quota,
\seats)$, $\winners$, $\losers$, $\bm{\tau}^R$, $\delta$): \\
1 \> $\winners_0$ $\leftarrow$ Subset of $\winners$ that have won on first preferences \\
2 \> $\assertion_{E,DL} \leftarrow \emptyset$ $\triangleright$ Our (eventual) dual-loop audit\\[5pt]
3 \> $\assertion_{\IQ} \leftarrow [\IQ(w) | \forall w \in \winners_0]$  $\triangleright$ Compute \underline{auditable} $\IQ$ assertions for winners in $\mathcal{W}_0$  \\
\\
4 \> \textbf{if} $\assertion_{\IQ} \equiv \emptyset$ \textbf{then return} $\emptyset$ $\triangleright$ A dual-loop audit is not possible \\[5pt]
5 \> \textbf{else if} $|\assertion_{\IQ}| \equiv |\winners|$ \textbf{then}  $\assertion_{E,DL} \leftarrow \assertion_{\IQ}$ $\triangleright$ Our $\IQ$ assertions form a full audit\\[5pt]
\> \textbf{else then} \\
6 \>\> $W' \leftarrow [w | \forall w \in \winners_0 \,.\, \IQ(w) \in \assertion_{\IQ}] $ \\
7 \>\> $\mathcal{R} \leftarrow \winners \setminus W'$  $\triangleright$ Compute set of unverified winners $\mathcal{R}$\\[5pt]
8 \>\> $\assertion_{full}, \assertion_{partial} \leftarrow \emptyset, \assertion_{\IQ}$  $\triangleright$ Initialise full and partial audits \\[5pt]
\>\> $\triangleright$ Initialise neighbourhood of transfer value lower bounds for candidates in $W'$ \\
9 \>\> $\bm{\underline{\tau}_0} \leftarrow [\tau^R_w - \delta | \forall w \in W']$ \\
10 \>\> $\bm{\underline{N}} \leftarrow [\bm{\underline{\tau}_0}]$ \\[5pt]

11 \>\> \textbf{while} $\bm{\underline{N}} \neq \emptyset$ \textbf{do} \\
\>\>\> $\triangleright$ Initialise `best' lower bound vector to the first in our neighbourhood\\
12 \>\>\> $\bm{\underline{\tau}}_{best} \leftarrow \bm{\underline{N}}[0]$  \\[5pt]
\>\>\> $\triangleright$ Consider each candidate transfer value lower bound vectors in our \\
\>\>\> neighbourhood, and run the \textsc{InnerLoop} (\autoref{alg:innerloop}) to find a full and \\
\>\>\> partial audits. Keep track of the best found full and partial audit. \\
13 \>\>\> \textbf{for} $\bm{\underline{\tau}} \in \bm{\underline{N}}$ \textbf{do} \\
14 \>\>\>\> $\assertion'_{full}, \assertion'_{partial}$ $\leftarrow$ \textsc{InnerLoop}($E$, $\winners$, $\bm{\tau}^R$, $\bm{\underline{\tau}}$, $\mathcal{R}$, $W'$, $\assertion_{\IQ}$, $\delta$) \\[5pt]
15 \>\>\>\> \textbf{if} $ASN(\assertion'_{full}) < ASN(\assertion_{full})$ \textbf{then} \\
16 \>\>\>\>\> $\assertion_{full}, \assertion_{partial}, \bm{\underline{\tau}}_{best}  \leftarrow \assertion'_{full}, \assertion'_{full}, \bm{\underline{\tau}}$ \\[5pt]
17 \>\>\>\> \textbf{else if} $\assertion'_{partial} \succ \assertion_{partial}$ \textbf{then} \\
18 \>\>\>\>\> $\assertion_{partial}, \bm{\underline{\tau}}_{best} \leftarrow \assertion'_{partial}, \bm{\underline{\tau}}$\\[5pt]
\>\>\> $\triangleright$ Find new candidate lower bound vectors  around $\bm{\underline{\tau}}_{best}$, if possible\\
19 \>\>\>  $\bm{\underline{N}} \leftarrow$ \textsc{Neighbours}$_{LB}$($\bm{\underline{\tau}}_{best}$)\\[5pt]
\>\>\> $\triangleright$ If the assertions used to validate the transfer value lower bounds are \\
\>\>\> not the most expensive of those in $\assertion_{partial}$ then continuing to explore \\
\>\>\> further lower bound vectors will not improve either $\assertion_{full}$ or $\assertion_{partial}$ \\
20 \>\>\> \textbf{if} $\LT$ assertions are not the most expensive in $\assertion_{partial}$ \textbf{then break}\\[5pt]
\>\> $\triangleright$ If we have found a full audit, return it as our dual-loop audit, otherwise \\
\>\> return the best partial audit. \\
21 \>\> \textbf{if} $\assertion_{full} \neq \emptyset$ \textbf{then} $\assertion_{E,DL} \leftarrow \assertion_{full}$ \textbf{else} $\assertion_{E,DL} \leftarrow \assertion_{partial}$ \\[5pt]
22 \> \textbf{return} $\assertion_{E,DL}$ 
\end{tabbing}
\caption{Outline of the outer-loop of the Dual-Loop STV Audit method for an STV election $\election = (\cand, \ballots, \quota,
\seats)$, with reported winners $\winners$, reported losers $\losers$, and reported transfer values $\tau^R_w$ for each winner $w \in \winners$,  $\bm{\tau}^R = [\tau^R_w | \forall w \in \winners]$. This outer-loop calls a procedure that takes a candidate transfer value lower bound vector and generates new candidates in its neighbourhood, \textsc{Neighbours}$_{LB}$.}
\label{fig:overviewdualloop}
\end{figure}

\begin{figure}[!t]
\centering
\begin{tabbing}
\hspace{0.25in}\=\hspace{0.25in}\=\hspace{0.25in}\=\hspace{0.25in}\=\hspace{0.25in}\=\hspace{0.25in}\=\kill
\textsc{ConstructAudit}($\election = (\cand, \ballots, \quota,
\seats)$, $\losers$, $\bm{\underline{\tau}}$, $\bm{\overline{\tau}}$, $W'$, $\assertion_{\IQ}$, $\mathcal{R}$): \\
1 \> $\assertion'_{full}, \assertion'_{partial} \leftarrow \emptyset, \assertion_{\IQ}$ $\triangleright$ Initialise a candidate full and partial audit \\[5pt]
\> $\triangleright$ Compute \underline{auditable} $\LT$ assertions to validate transfer value lower bounds in $\bm{\underline{\tau}}$ \\
2 \> $\assertion_{\LT} \leftarrow [\LT(w, \bm{\underline{\tau}}_w) | \forall w \in W']$ \\[5pt]
\> $\triangleright$ If we cannot verify all  lower bounds in $\bm{\underline{\tau}}$, then we cannot form a full audit and \\ 
\> our partial audit verifies only those winners for which we formed $\IQ$ assertions. \\
3 \> \textbf{if} $|\assertion_{\LT}| < |W'|$ \textbf{then return } $\emptyset, \assertion'_{partial}$ \\[5pt]
\> $\triangleright$ Compute auditable $\UT$ assertions to validate transfer value upper bounds in $\bm{\overline{\tau}}$ \\
4 \> $\assertion_{\UT} \leftarrow [\UT(w, \bm{\overline{\tau}}_w) | \forall w \in W']$ \\[5pt]
\> $\triangleright$ If we cannot verify all  upper bounds in $\bm{\overline{\tau}}$, then we cannot form a full audit and \\
\> our partial audit verifies only those winners for which we formed $\IQ$ assertions. \\
5 \> \textbf{if} $|\assertion_{\UT}| < |W'|$ \textbf{then return } $\emptyset, \assertion'_{partial}$ \\[5pt]
\> $\triangleright$ Consider three approaches for verifying each unverified winner $r \in \mathcal{R}$ \\
6 \> $\assertion_{\mathcal{R}} \leftarrow \emptyset$, Verified $\leftarrow$ 0  \\
7 \> \textbf{for each} $r \in \mathcal{R}$ \textbf{do} \\
8 \> \> $\assertion_{\AGS}, \assertion_{Vo1} \leftarrow $ \textsc{AuditByAG}$^*$($E$, $\losers$, $W'$, $\bm{\underline{\tau}},$ $\bm{\overline{\tau}}$, $r$) \\
9 \> \> $\assertion_{Vo2} \leftarrow $ \textsc{AuditByNL}$^*$($E$, $\losers$, $W'$, $\bm{\underline{\tau}},$ $\bm{\overline{\tau}}$, $r$, $\assertion_{\AGS}$) \\
10 \>\> $\assertion_{Vo3} \leftarrow $ \textsc{AuditByIQX}($E$,  $W'$, $\bm{\underline{\tau}},$ $\bm{\overline{\tau}}$, $r$, $\assertion_{\AGS}$) \\
11 \>\> \textbf{if} at least one $\assertion_{Voi} \neq \emptyset$ \textbf{then} \\
12 \>\>\> $\assertion_{r} \leftarrow $ Cheapest of the assertion sets $\assertion_{Vo1}$, $\assertion_{Vo2}$, and $\assertion_{Vo3}$ \\
13 \>\>\> $\assertion_\mathcal{R} \leftarrow \assertion_{\mathcal{R}} \cup \assertion_{r}$ \\
14 \>\>\> Verified $\leftarrow $ Verified + 1\\[5pt]
15\> \textbf{if} Verified $>$ 0 \textbf{then} \\
16\>\> $\assertion'_{partial} \leftarrow \assertion_{\IQ} \cup \assertion_{\LT} \cup \assertion_{\UT} \cup \assertion_{\mathcal{R}}$ \\
17\>\> \textbf{if} Verified $\equiv$ $|\mathcal{R}|$ \textbf{then}  $\assertion'_{full} \leftarrow \assertion'_{partial}$ \\[5pt]
18 \> \textbf{return} $\assertion'_{full}, \assertion'_{partial}$ 
\end{tabbing}
\caption{Construction of a candidate full and/or partial audit for an STV election $\election = (\cand, \ballots, \quota,
\seats)$ given: reported losers $\losers$; assumed winners on first preferences $W'$; assumed lower and upper bounds on the transfer values of candidates in $W'$, $\bm{\underline{\tau}}$ and $\bm{\overline{\tau}}$; assertions used to verify that the candidates in $W'$ won on first preferences, $\assertion_{\IQ}$; and the set of currently unverified winners, $\mathcal{R}$. \vspace{-10pt}} 
\label{fig:constructaudit}
\end{figure}

\vspace{-10pt}
\section{Dual-Loop Audits: The Details}\label{sec:Detail}

In this section, we describe the three core components of the dual-loop method in more detail: the outer loop; the inner loop; and the procedure for constructing an audit given a candidate transfer value lower and upper bound vector. 

\mysubsection{Constructing an Audit} 
\autoref{fig:constructaudit} outlines the steps performed when constructing a full and partial audit given: a set of winners assumed to have won on first preferences, $W'$; a vector of lower and upper bounds on transfer values for each candidate in $W'$, $\bm{\underline{\tau}}$  and $\bm{\overline{\tau}}$; a set of assertions verifying that candidates $W'$ won on first preferences, $\assertion_{\IQ}$; and reported winners $\mathcal{R}$ that need to be verified. 

\textsc{ConstructAudit} first initialises its candidate full and partial audit to an empty set and the set of auditable $\IQ$ assertions, $\assertion_{\IQ}$, respectively (step 1).
It then attempts to verify the set of transfer value lower bounds, $\bm{\underline{\tau}}$, by forming all of the auditable $\LT$ assertions it can form for each $w \in W'$ (step 2), $\assertion_{\LT}$.
If $|\assertion_{\LT}| < |W'|$, then we can't validate the given set of transfer value lower bounds, and return the empty $\assertion'_{full}$ and unchanged $\assertion'_{partial}$ (step 3).
We follow the same process to verify the set of transfer value upper bounds, $\bm{\overline{\tau}}$,  in steps 4--5. 

Now that we have validated a set of lower and upper bounds on first winner transfer values, we use these bounds to help us verify our other winners $\mathcal{R}$. For a given $r \in \mathcal{R}$, we consider three ways of verifying them. Let $\assertion_{Voi}$ denote the set of assertions formed by $i$'th verification approach.\\

\noindent \textbf{Vo1 (Step 8)} We compute all possible auditable $\AGS$ relationships between reported winner $r$ and each reported loser $l \in \losers$.
    \[
    \assertion_{\AGS} \leftarrow [\AGS(r, l, W', \bm{\underline{\tau}}, \bm{\overline{\tau}}) | \forall l \in \losers]
    \]
    If $|\assertion_{\AGS}| \equiv |\losers|$, we verify $r$ with  $\assertion_{Vo1} =  \assertion_{\AGS}$. Otherwise, $\assertion_{Vo1} = \emptyset$.\\

\noindent \textbf{Vo2 (Step 9)} We compute all possible auditable $\CNEBS$ relationships between reported winner $r$ and each reported loser $l \in \losers$, reusing  $\assertion_{\AGS}$.
    \[
    \assertion_{\CNEBS} \leftarrow [\CNEBS(r, l, W', \bm{\underline{\tau}}, \bm{\overline{\tau}}, O^*) | \forall l \in \losers]
    \]
    Here, $O^*$ denotes the set of candidates $o$ for which $\AGS(r, o, W', \bm{\underline{\tau}}, \bm{\overline{\tau}}) \in \assertion_{\AGS}$.
    If $|\assertion_{\CNEBS}| \equiv |\losers|$, we verify $r$ as a winner with assertions $\assertion_{Vo2} = \assertion_{\CNEBS} \cup \assertion_{\AGS}$. Otherwise, if we have been able to form \textit{some} $\CNEBS$ assertions between $r$ and a subset of the losers $L \subset \losers$, we may be able to use these to help us form $\CNEBS$ assertions between $r$ and the losers in $\losers \setminus L$. We  iterate this verification method until we reach a fixed point. The first iteration of this loop forms the assertions:
        \[
    \assertion_{\CNEBS,0} \leftarrow [\CNEBS(r, l, W', \bm{\underline{\tau}}, \bm{\overline{\tau}}, O^*) | \forall l \in \losers]
    \]

    Let $L_0$ denote the set of losers $l$ for which  $\CNEBS(r, l, \ldots) \in  \assertion_{\CNEBS,0}$. 
    The $i^{th}$ iteration of the loop forms the assertions:
            \[
     \assertion_{\CNEBS,i} \leftarrow  \assertion_{\CNEBS,{i-1}} \cup [\CNEBS(r, l, W', \bm{\underline{\tau}}, \bm{\overline{\tau}},  O^*_i) | \forall l \in \losers \setminus L_i]
    \]
 Here, $O^*_i$ denotes the set of candidates $o$ for which \textit{either} $\AGS(r, o, W', \bm{\underline{\tau}}, $ $\bm{\overline{\tau}}) \in \assertion_{\AGS}$ or $\CNEBS(r, o, W', \bm{\underline{\tau}}, \bm{\overline{\tau}, \_}) \in \assertion_{\CNEBS,{i-1}}$.\\
 
    We terminate this process when $\assertion_{\CNEBS,i} \equiv \assertion_{\CNEBS,i-1}$. If $|\assertion_{\CNEBS,i}| \equiv |\losers|$, we verify $r$ as a winner with assertions $\assertion_{Vo2} = \assertion_{\CNEBS,i} \cup \assertion_{\AGS}$. Otherwise, $\assertion_{Vo2} = \emptyset$.\\

\noindent \textbf{Vo3 (Step 10)} If $a = \IQX(r, W', \bm{\underline{\tau}}, \bm{\overline{\tau}}, O^*)$, where $O^*$ is, as previously, the set of candidates $o$ for which $\AGS(r, o, \bm{\underline{\tau}}, \bm{\overline{\tau}}) \in \assertion_{\AGS}$, can be formed and is auditable, we verify $r$ as a winner with assertions $\assertion_{Vo3} = \{a\} \cup \assertion_{\AGS}$. 
\vspace{3pt}

If none of these methods are successful, we move on to consider the next unverified winner in $r$. Otherwise, we use the set of assertions $\assertion_{Voi}$ with the smallest ASN to verify winner $r$ (step 12). For all $r \in \mathcal{R}$ that we can verify using one of these three verification methods, we combine each $\assertion_r$ to form the set $\assertion_{\mathcal{R}}$ (step 13). 
If $\assertion_{\mathcal{R}} \neq \emptyset$, we form a new partial audit (step 16).
If we have verified all winners in $\mathcal{R}$, this audit is a full RLA, and we set  $\assertion'_{full} \equiv \assertion'_{partial}$ (step 17). 
We return, for the given lower and upper bounds on first winner transfer values, $\bm{\underline{\tau}}$  and $\bm{\overline{\tau}}$, these full and partial audits  in step 18.

Note that we do not need to form an $\LT$ assertion for a candidate if the assumed lower bound on their transfer value is 0, or a $\UT$ assertion  when the assumed transfer value upper bound  is the theoretical maximum, $\tau_{max}$.

\begin{figure}[!t]
\centering
\begin{tabbing}
\hspace{0.25in}\=\hspace{0.25in}\=\hspace{0.25in}\=\hspace{0.25in}\=\hspace{0.25in}\=\hspace{0.25in}\=\kill
\textsc{InnerLoop}($\election = (\cand, \ballots, \quota,
\seats)$, $\winners$, $\losers$, $\bm{\tau}^R$, $\bm{\underline{\tau}}$, $\mathcal{R}$, $W'$, $\assertion_{\IQ}$, $\delta$): \\
1 \> $\assertion'_{full}, \assertion'_{partial} \leftarrow \emptyset, \assertion_{\IQ}$   $\triangleright$ Initialise full and partial audits \\[5pt]
\> $\triangleright$ Initialise neighbourhood of transfer value upper bounds for candidates in $W'$ \\
2 \> $\bm{\overline{\tau}_0} \leftarrow [\tau^R_w + \delta | \forall w \in W']$ \\
3 \> $\bm{\overline{N}} \leftarrow [\bm{\overline{\tau}_0}]$ \\[5pt]
4 \> \textbf{while} $\bm{\overline{N}} \neq \emptyset$ \textbf{do} \\
\>\> $\triangleright$ Initialise `best' upper bound vector to the first in our neighbourhood\\
5 \>\> $\bm{\overline{\tau}}_{best} \leftarrow \bm{\overline{N}}[0]$  \\[5pt]
\>\> $\triangleright$ Consider each candidate transfer value upper bound vector in our \\
\>\> neighbourhood, and run  \textsc{ConstructAudit} (\autoref{fig:constructaudit}) to find a full and \\
\>\> partial audits. Keep track of the best found full and partial audit. \\
6 \>\> \textbf{for} $\bm{\overline{\tau}} \in \bm{\overline{N}}$ \textbf{do} \\
7 \>\>\> $\assertion''_{full}, \assertion''_{partial}$ $\leftarrow$ \textsc{ConstructAudit}($\election$, $\losers$, $\bm{\underline{\tau}}$, $\bm{\overline{\tau}}$, $W'$, $\assertion_{\IQ}$, $\mathcal{R}$) \\[5pt]
8 \>\>\> \textbf{if} $ASN(\assertion''_{full}) < ASN(\assertion'_{full})$ \textbf{then} \\
9 \>\>\>\> $\assertion'_{full}, \assertion'_{partial}, \bm{\overline{\tau}}_{best}  \leftarrow \assertion''_{full}, \assertion''_{full}, \bm{\overline{\tau}}$ \\[5pt]
10 \>\>\> \textbf{else if} $\assertion''_{partial} \succ \assertion'_{partial}$ \textbf{then} \\
11 \>\>\>\> $\assertion'_{partial}, \bm{\overline{\tau}}_{best} \leftarrow \assertion''_{partial}, \bm{\overline{\tau}}$\\[5pt]
\>\> $\triangleright$ Find new candidate upper bound vectors  around $\bm{\overline{\tau}}_{best}$, if possible\\
12 \>\> \textbf{if} $\bm{\overline{\tau}}_{best} \neq \emptyset$ \textbf{then}  $\bm{\overline{N}} \leftarrow$ \textsc{Neighbours}$_{UB}$($\bm{\overline{\tau}}_{best}$) \textbf{else} $\bm{\overline{N}} \leftarrow \emptyset$\\[5pt]
\>\> $\triangleright$ If the assertions used to validate the transfer value upper bounds are \\
\>\> not the most expensive of those in $\assertion'_{partial}$ then continuing to explore \\
\>\> further upper bound vectors will not improve either $\assertion'_{full}$ or $\assertion'_{partial}$ \\
13 \>\> \textbf{if} $\UT$ assertions are not the most expensive in $\assertion'_{partial}$ \textbf{then break}\\[5pt]
14 \> \textbf{return} $\assertion'_{full}$, $\assertion'_{partial}$ 
\end{tabbing}
\caption{Inner-loop of the Dual-Loop audit method for an STV election $\election = (\cand, \ballots, \quota,
\seats)$, with: reported winners and losers $\winners$, $\losers$; reported transfer values $\tau^R_w$ for each $w \in \winners$,  $\bm{\tau}^R = [\tau^R_w | \forall w \in \winners]$; assumed winners on first preferences $W'$; assumed lower  bounds on transfer values for $W'$, $\bm{\underline{\tau}}$; assertions used to verify that the candidates in $W'$ won on first preferences, $\assertion_{\IQ}$; and the set of  unverified winners, $\mathcal{R}$. The inner-loop calls a procedure that takes a candidate transfer value upper bound vector and generates new candidates, \textsc{Neighbours}$_{UB}$. \vspace{-15pt}}
\label{alg:innerloop}
\end{figure}

\mysubsection{Inner Loop} 
Given a vector of lower bounds on verified first winner transfer values, $\bm{\underline{\tau}}$, the inner loop will start with a neighbourhood $\overline{N}$ containing just one upper bound vector, $\overline{N}$ $=$ [$\bm{\overline{\tau}}_0$], where the upper bound for winner $w \in W'$ in $\bm{\overline{\tau}}_0$ is set to their reported transfer value plus $\delta \ll 1$. 
For each neighbourhood $\overline{N}$ that the inner loop considers, each upper bound vector in $\overline{N}$, $\bm{\overline{\tau}}_i$, is considered in turn. The \textsc{ConstructAudit} procedure is called for the pair of  vectors $\bm{\underline{\tau}}$ and $\bm{\overline{\tau}}_i$, finding a candidate full and partial audit, $\assertion'_{full}$ and $\assertion'_{partial}$. 

\textit{Updating our running full and partial audits.} If $ASN(\assertion'_{full}) < ASN(\assertion_{full})$, then we have found a better full audit and we replace  $\assertion_{full}$ and $\assertion_{partial}$ with $\assertion'_{full}$. 
If $\assertion'_{full} = \emptyset$, and $\assertion'_{partial}$ either (i)  verifies \textit{more} winners than $\assertion_{partial}$, or (ii) verifies the same number of winners but at a cheaper cost, $ASN(\assertion'_{partial}) < ASN(\assertion_{partial})$, then we replace the current $\assertion_{partial}$ with $\assertion'_{partial}$.

\textit{Terminating the Inner Loop.} Recall that the inner loop searches for audits or improved audits by iterating over possible upper bound transfer value vectors for candidates seated on their first preferences. These candidate upper bounds are gradually increased over the course of the loop. Increasing these upper bounds makes our $\UT$ assertions easier to form (with smaller ASNs), or possible to form, but makes verifying each reported winner in $\mathcal{R}$ with $\AGS$, $\CNEBS$, and $\IQX$ assertions more difficult. If we can form all required $\UT$ assertions, and their cost is not the dominant cost in our best partial audit, then we can break out of the inner loop. Searching further will only result in cheaper $\UT$ assertions at the expense of increasing the cost of the $\AGS$, $\CNEBS$, and $\IQX$ needed to verify our winners. 

\textit{Generating a new neighbourhood.} If, after considering each upper bound vector in $\overline{N}$, our termination condition has not been satisfied,   we take the upper bound vector used to form the current best full or partial audit, or the first upper bound vector in $\overline{N}$, $\bm{\overline{\tau}}$, and generate a new neighbourhood of upper bound vectors. This neighbourhood consists of up to $|W'|$ new vectors, one for each $w \in W'$ where $\bm{\overline{\tau}}_w < \tau_{max}$, where we replace $\bm{\overline{\tau}}_w$ with $\min( \tau_{max}, \bm{\overline{\tau}}_w + \delta)$.  The inner loop then explores this new neighbourhood as described above. The inner loop also terminates when we cannot form a new neighbourhood, i.e., all transfer value upper bounds are at their maximum theoretical value $\tau_{max}$.

\mysubsection{Outer Loop} 
Steps 9--21 of \autoref{fig:overviewdualloop} represent the outer loop of the dual-loop method. The outer loop starts with a neighbourhood of candidate transfer value lower bounds for winners on first preferences, $\bm{\underline{N}}$, initialised to [$\bm{\underline{\tau}}_0$], where the lower bound for winner $w \in W'$ in $\bm{\underline{\tau}}_0$ is set to their reported transfer value minus $\delta \ll 1$ (steps 9--10). The outer loop considers each candidate  lower bound vector in $\bm{\underline{N}}$, and executes the inner loop (steps 13--18). The inner loop returns a candidate full and partial audit for a given lower bound vector (step 14). We keep track of the best found full and partial audit as we explore  candidate lower bound vectors (steps 15--18), and the `best' transfer value lower bound vector in the current neighbourhood,  $\bm{\underline{\tau}}_{best}$ (steps 16 and 18). Once we have explored the current neighbourhood, we generate a new one, if possible, from $\bm{\underline{\tau}}_{best}$ (step 19).    

 \textit{Terminating the Outer Loop.} Candidate transfer value lower bounds are reduced over the course of the outer loop. Decreasing these bounds makes our $\LT$ assertions easier to form (with smaller ASNs), but makes verifying each reported winner in $\mathcal{R}$ with $\AGS$, $\CNEBS$, and $\IQX$ assertions more difficult. If we can form all required $\LT$ assertions, and their cost is not the dominant cost in our best partial audit, then we can break out of the outer loop (step 20). Searching further will only result in cheaper $\LT$ assertions at the expense of increasing the cost of the $\AGS$, $\CNEBS$, and $\IQX$ needed to verify our winners.

\textit{Generating a new neighbourhood (\textsc{Neighbours}$_{LB}$).} If, after considering each lower bound vector in $\bm{\underline{N}}$, our termination condition is unsatisfied,   we take one of the  vectors in our neighbourhood, $\bm{\underline{\tau}}_{best}$, and generate a new neighbourhood of lower bound vectors (step 19).  The new neighbourhood consists of up to $|W'|$ new vectors, one for each $w \in W'$ where $\bm{\underline{\tau}}_w > 0$, in which we replace $\bm{\underline{\tau}}_w$ with $\bm{\underline{\tau}}_w - \delta$, if $\bm{\underline{\tau}}_w > \delta$, and 0 otherwise. The outer loop  terminates if we cannot form a new neighbourhood, i.e., all transfer value lower bounds in $\bm{\underline{\tau}}_{best}$ are 0.

\section{Results}

\begin{table}[!t]
\centering\footnotesize
\begin{tabular}{c|c|c|c|c}
\toprule
Winners  & \multicolumn{2}{c|}{3 seats} & \multicolumn{2}{c}{4 seats} \\
\cline{2-5}
verified & Instances (/252) & ASN Avg (Min, Max) & Instances (/261) & ASN Avg (Min, Max)  \\
\toprule
None & 5 (2\%) &  &  4 (2\%) &  \\
1 & 28  (11\%) & 74 (14, 289)  & 25 (10\%) &  98 (16, 457) \\
2 & 73  (29\%) & 281 (27, 2036)  & 69 (26\%) & 331 (36, 1677) \\
3 & 146 (58\%) & 379 (35, 1790)  & 79 (30\%) & 406 (54, 2033) \\
4 &    &  & 84  (32\%) &  635 (71, 2339) \\
\bottomrule
\end{tabular}
\caption{Across  252 3-seat and 261 4-seat STV contests, we report: the number and percentage of instances where we verify no  or only 1 to 4 winners within an ASN of 2500 ballots; and the average and range of ASNs for these audits. \vspace{-20pt}}
\label{tab:SummaryResults}
\end{table}

\begin{figure}[t]
\centering
\includegraphics[width=\textwidth]{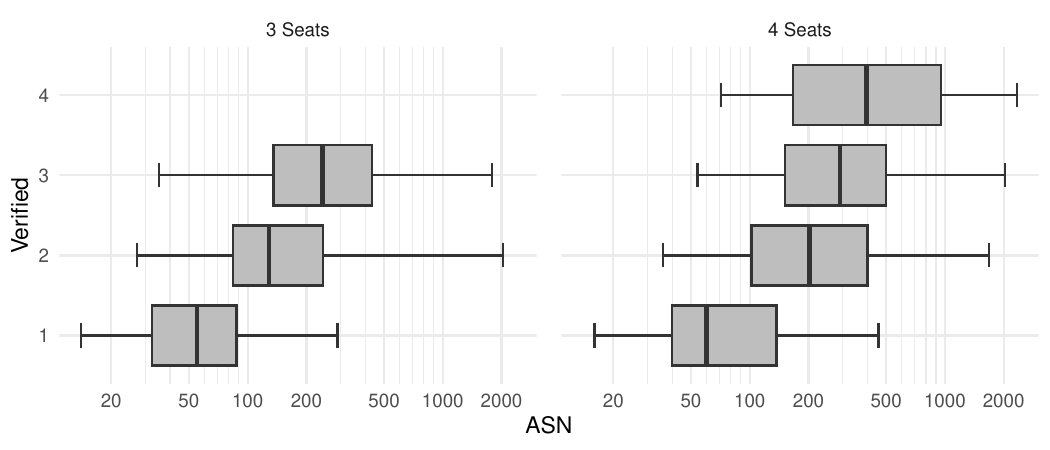}
\caption{Side-by-side box plots showing the spread of ASNs (x-axes in log scale) of \textit{formed} audits that verify varying numbers of winners for our data set of 252 3-seat (left) and 261 4-seat (right) STV contests. For each box plot: the vertical line in the middle of each `box' shows the median ASN for each group, the edges of the boxes are the first and third quartiles, and the `whiskers' extending out of each box show the minimum and maximum ASN. \vspace{-10pt}}
\label{fig:ASNBoxPlot}
\end{figure}

We used data from 513 three- or four-seat STV elections from local council elections in Scotland in 2017 and 2022. Scotland uses the Weighted Inclusive Gregory method at the local council level. Each election in this data set satisfies the `first winner' criterion, and involves 3 to 13 candidates.\footnote{It is not possible to say how often the first winner criterion would be satisfied by STV elections in general, however it becomes more likely with an increased number of seats, smaller quotas, and the presence of at least one dominant candidate.} For each of these contests, we applied the method described in \autoref{sec:Overview} and \autoref{sec:Detail} to form a set of assertions, designed to verify as many reported winners as possible.      

We computed ASN estimates via simulation for each assertion. The simulation considered incrementally larger sample sizes, randomly generating 1-vote overstatements (errors that overstate the margin of the assertion by one) at a defined rate, until the risk value of the audit fell below the desired limit. This simulation was repeated, and the average sample size needed to meet the risk limit, rounded up, formed the ASN for the assertion. For sample size estimations, we used a risk limit of 5\%,  an expected error rate of 2 overstatements per 1000 ballots, 20 simulations, $\delta = 0.005$, and the ALPHA risk function  \cite{shangrlaSHORT}. 

Across the 252 3-seat and 261 4-seat contests in our data set, \autoref{tab:SummaryResults} reports the number, and percentage, of instances in which we \textit{can form an audit} that verifies a specific number of winners, from 0 to 4, alongside the average, minimum, and maximum ASNs for these audits. For this paper, we considered a contest to be auditable if its ASN was $M$ = 2500 ballots or less. The average number of ballots cast in the Scottish elections in our dataset is just under 5500. Our choice of $M$ in this case was designed to give us a sense of how many elections we could audit with a reasonably modest sample size. In place of a fixed $M$, we could alternately set $M$ to a defined percentage of the number of ballots cast.   \autoref{fig:ASNBoxPlot} presents two side-by-side box plots visualising the distribution of ASNs across audits that verify 1 to 3 winners in 3-seat contests (left) and 1 to 4 winners in 4-seat contests (right). All code used to form these audits can be found at: \url{https://github.com/michelleblom/stv-rla}.

\section{Conclusion}
Finding RLAs for STV elections is challenging, since the election process is complex and ballots change in value during tabulation. 
Previous methods were limited to STV elections of  2 candidates. In this paper we extend this to ``arbitrarily'' large elections, as long as some candidates obtained a quota in the first round.  While this allows us to, at least partially, audit many elections, there remains much further work. To date, we have been most successful in tackling STV elections with a relatively small number of candidates. On the Scottish dataset, our ability to verify all winners reduces as the number of seats increases. In general, elections with more seats involve more candidates, and more surplus transfers, which can make them harder to audit with current methods. The heuristic presented in this paper could be improved by taking a more nuanced view when forming  partial audits by identifying the relative cost of auditing each winner.    

\subsubsection{Acknowledgements.}

This work was supported by the Australian Research Council
(Discovery Project DP220101012, OPTIMA ITTC IC200100009).

\bibliographystyle{splncs04}
\bibliography{refs}

\end{document}